\documentclass[12pt,a4paper]{article}%

\usepackage{amsmath,amssymb,amsfonts}
\usepackage{caption2}
\usepackage[]{hyperref}
\usepackage[pdftex]{color,graphicx}

\numberwithin{equation}{section} 
\setcaptionmargin{1cm}

\begin{document}
\begin{titlepage}
\vskip 1.0cm
\begin{center}
{\Large \bf On the QCD corrections to $\Delta F=2$ FCNC in the Supersymmetric SM with hierarchical squark masses }
\vskip 1.0cm {\large  Enrico Bertuzzo, Marco Farina and Paolo Lodone} \\[1cm]
{\it Scuola Normale Superiore and INFN, Piazza dei Cavalieri 7, 56126 Pisa, Italy} \\[5mm]
\vskip 1.0cm
\today
\end{center}

\begin{abstract}
In the context of the Supersymmetric Standard Model with hierarchical sfermion masses, we re-analyze the QCD corrections to $\Delta F=2$ effective Hamiltonian properly including effects which have been neglected so far.
The point is that some $\Delta F=2$ diagrams, involving both the heavy and the light sparticles, exhibit a logarithmic dependence on the ratio between the two masses, signalling a sensitivity to all the momenta between the two scales.
In order to properly deal with these terms one has to take into account the mixing between $\Delta F=1$ and $\Delta F=2$ operators.
In typical situations this treatment can affect the result at the level of $30\%$ or even more.
\end{abstract}
\end{titlepage}

\section{Introduction }

The experimental data unambiguosly tell us that the flavour structure of any model for New Physics at the TeV scale must be highly non-generic.
In the context of low energy Supersymmetry, after imposing the bounds from Flavour Changing Neutral Currents (FCNC) the allowed portion of parameter space is so small that one usually refers to this fact as the Supersymmetric Flavour Problem.
Among the various investigations which try to solve or at least alleviate the problem, many motivated proposals suggest that the (s)particle spectrum may be hierarchical, with the sfermions of the first two generations much heavier than those of the third one \cite{Dine:1993np}-\cite{Giudice:2008uk}.

Very stringent bounds come from $\Delta F=2$ processes involved in $K$ and $B$ physics (for a recent review see \cite{Isidori:2010kg}).
For this reason, in order to make a comparison with data, it is important to resum the QCD corrections with the usual means of effective field theories. These corrections have already been computed \cite{Bagger:1997gg}-\cite{Contino:1998nw}, however there is an effect which has been so far neglected although it can be important in the case of a large hierarchy.

If one naively computes the relevant diagrams at leading order, then those involving both the heavy and the light squarks\footnote{Notice that there can be situations in which this contribution is the dominant one, see for example \cite{Barbieri:2010ar}.} produce terms proportional to the logarithm of the ratio of the two masses (we assume that the various gauginos have typical mass close to that of the third generation). This is a clear signal that the diagram is sensitive to all the momenta between the two scales and thus that these logarithms cannot be considered as initial conditions for the Wilson coefficient at a given scale.
On the other hand the diagrams involving only the heavy (light) squarks are only sensitive to the higher (lower) mass scale, and they do not exhibit such logarithms so that they can be treated in the standard way.

This situation is somewhat analogous to what happens in the Standard Model when one has to deal with logarithms of $m_W/m_c$ \cite{Herrlich:1996vf}. In that case in order to resum these logarithms to all orders in perturbation theory one has to consider the RG evolution, from $m_W$ down to $m_c$, for the coefficients of all the relevant operators in the effective theory in which the $W$ boson and the top quark are integrated out. Then at the charm threshold one can do the matching with the theory in which $c$ is integrated out too, so that at the end of the day one obtains RG improved Wilson coefficients that multiply matrix elements which do not contain large logarithms.

Analogously in this context $\Delta F=2$ and several $\Delta F=1$ operators are generated at the high scale, which then mix through the exchange of third generation squarks. This produces final results that, at lowest order in the coupling constant, are equal to the naive logarithms one would find directly from the original diagrams. The difference is that now the QCD corrections have been properly included at leading order, with resummed large logarithms.
A first step in this direction has already been made in \cite{Barbieri:2010ar}, in which only the leading gluino diagrams are properly accounted for. In this paper we extend this treatment also to the other contributions involving weak interactions, for the case of $K$ physics.

This work is structured as follows: in Section \ref{sect:struct} we specify our setup and recall basic facts about the amplitudes that we consider, in Section \ref{sect:mix} we study the flavour-violating mixing between the relevant operators, and finally in Section \ref{sect:evol} we present and discuss the results.

\section{Structure of the heavy-light contributions to FCNC} \label{sect:struct} 

In order to avoid unnecessary complications and focus on our issue, we consider the following setup:
\begin{enumerate}
\item There is a hierarchy between the mass of the (heavy) squarks of the first two generations $m_1, \, m_2 \sim m_h$ and that of the third one and
the other light sparticles $m_3, \, m_{\tilde{g}}, \, m_{\chi} \sim m_\ell $;
\item We are only interested in the $\Delta F=2$ contributions coming from the simultaneous exchange of both heavy and light squarks. For this reason
we can totally neglect the Higgs sector, which couples mainly to the third generation;
\item We neglect the mixing between LH and RH squarks as well as between gauginos;

\item We ignore the part of the lagrangian involving right handed down quarks. They can be included with the same procedure we outline below.

\item We focus on the flavour violation in the down quark sector.

\end{enumerate}

In conclusion, neglecting terms suppressed by the Yukawa coupling of down-type quarks, we focus on the Lagrangian:
\begin{eqnarray}
\mathcal{L}_{FC} &=& 
- g\, \tilde{u}^*_L T \,\overline{\tilde{W}^- }\, d_L \,+ \frac{g}{\sqrt{2}}\, \tilde{d}_L^* V \, \overline{\tilde{W}^3}\, d_L \nonumber \\
&& -\sqrt{2} \frac{g'}{6} \tilde{d}^*_L\, V \,\overline{\tilde{B}}\, d_L -\sqrt{2}\, g_3 \, \tilde{d}_L^* \, \lambda^b\, V\, \overline{\tilde{g}^b }\,d_L
\label{LFC}
\end{eqnarray}
in which all the fields are mass eigenstates, and $V,\, T$ are unitary mixing matrices. 
The contribution to the effective lagrangian for $\Delta F=2$ processes due to the exchange of one heavy and one light squark is:
\begin{equation} \label{eq:L123}
\mathcal{L}^{\Delta F=2}_{h \ell} = S_{h \ell} \, Q_1 + h.c.
\end{equation}
where $Q_1$ is the $\Delta F = 2$ operator:
\begin{equation}
Q_1 = (\overline{d}^\rho \gamma^\mu P_L {s}^\rho) \, (\overline{d}^\sigma \gamma_\mu P_L {s}^\sigma)
\label{Q1}
\end{equation}
with $\rho, \sigma$ colour indices and we focus on the $\Delta S =2$ case.
For later convenience we define:
$$
\xi_{j}^{\alpha\beta}=V_{j\alpha} V^*_{j\beta}\, ,~~~~~~~\eta_{j}^{\alpha\beta}=T_{j\alpha} T^*_{j\beta}
$$
where the roman index refers to the squarks and the greek indices to the quarks, that we are going to drop from now on.

Naively computing the various box diagrams with one heavy squark with mass $m_h$ and one light squark with mass $m_\ell$ one obtains the QCD-uncorrected amplitude (see e.g. \cite{Bertolini:1990if}\cite{Gabbiani:1996hi}):
\begin{equation}
S_{h \ell} =
   \frac{\xi_h \xi_\ell}{m_h^2}\left[ \alpha_s^2 \left(-\frac{37}{36}+ \frac{11}{18}\log{\frac{m_h^2}{m_{\ell}^2}}\right)
+\alpha_w^2\left(\frac{1}{4}\frac{\eta_h \eta_\ell}{\xi_h \xi_\ell} +  \left(\frac{1}{4}\frac{\eta_h \eta_\ell}{\xi_h \xi_\ell}+ \frac{2}{3}\frac{\alpha_s}{\alpha_w} \mathcal{R}+\mathcal{R}^2\right)\left( -\frac{5}{2}+\log{\frac{m_h^2}{m_{\ell}^2}}\right)\right)\right] \,
 \label{eq:naive}
 \end{equation}
where:
\begin{equation}
\mathcal{R}= \frac{1}{4} + \frac{1}{36} t_W^2 \, , \, \, \, t_W = \tan \theta_W \, .
\end{equation}
We immediatly recognize the feature stressed in the Introduction. In the case of large separation between $m_h$ and $m_\ell$, it is not consistent to use these expressions as initial condition for the coefficient of $Q_1$ at the scale $m_h$.
We thus need a more careful treatment of the QCD running between the two scales, as we now discuss.

\section{Mixing between $\Delta$F=2 and $\Delta$F=1 operators}  \label{sect:mix}

\begin{figure}[t]
\begin{center}
\begin{tabular}{cc}
\includegraphics[width=0.26\textwidth]{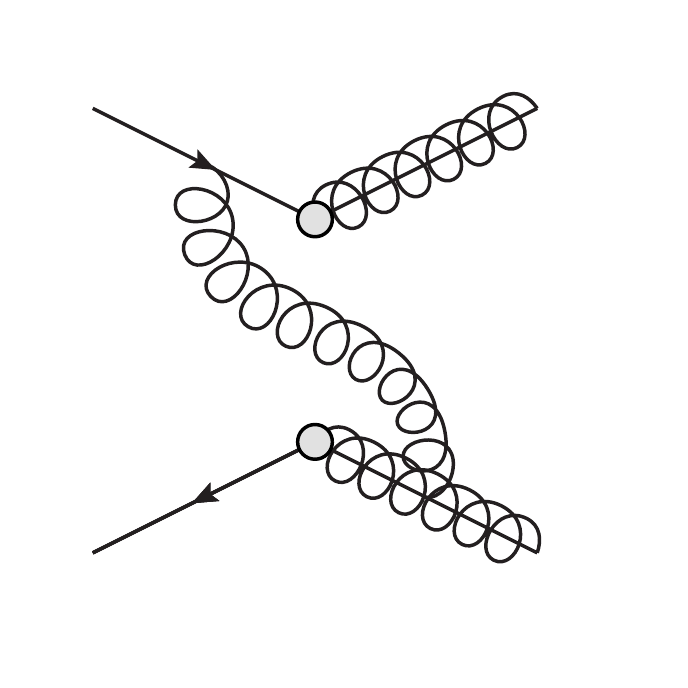} &
\includegraphics[width=0.35\textwidth]{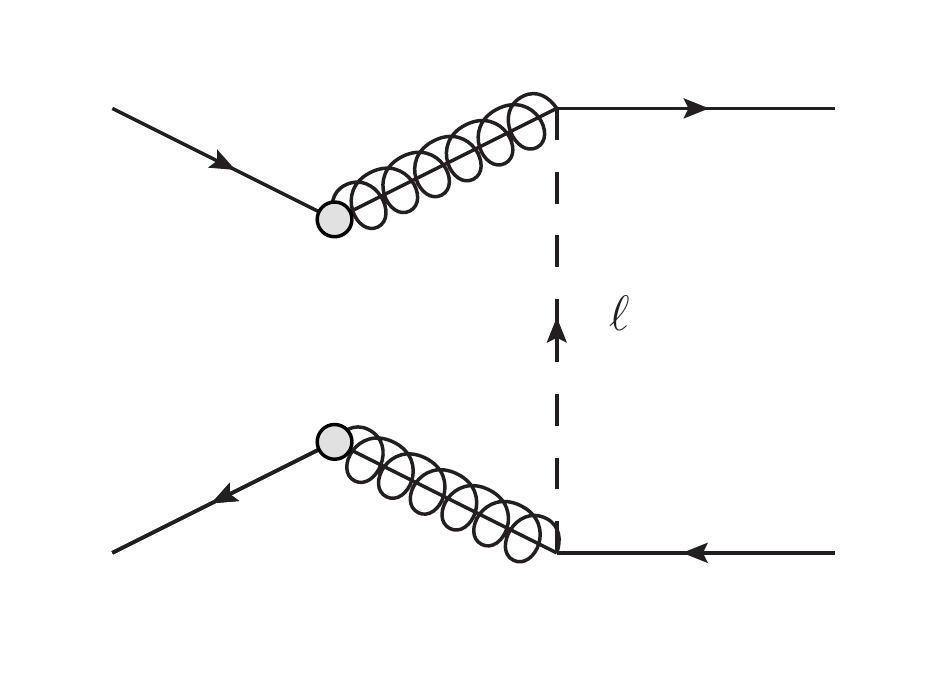}
\end{tabular}
\end{center}
\caption{\small{\it Examples of diagrams contributing to the QCD running: flavour conserving gluon exchange (left panel), and flavour violating exchange of a light squark (right panel).}}
\label{figure:Diagrams}
\end{figure}

The new ingredient that is required in order to deal with the heavy-light exchange in $\mathcal{L}^{\Delta F =2}_{h\ell}$ is the mixing between $Q_1$,
given by (\ref{Q1}), and the $\Delta F = 1$ operators which are generated after integrating out the heavy squarks of the first two generations (see Figure \ref{figure:Diagrams}).
A possible basis for these operators, according to the external light particles, is:
\begin{itemize}
 \item two gluinos:
\begin{equation}
\begin{array}{lll}
Q_1^g &=&\delta^{ab}\delta_{\beta\alpha}(\overline{d}^\beta P_R \widetilde{g}^b)(\overline{\widetilde{g}^a}P_L s^\alpha )\\
Q_2^g &=& d^{bac} t^c_{\beta\alpha}(\overline{d}^\beta P_R \widetilde{g}^b)(\overline{\widetilde{g}^a}P_L s^\alpha )\\
Q_3^g &=&i f^{bac} t^c_{\beta\alpha}(\overline{d}^\beta P_R \widetilde{g}^b)(\overline{\widetilde{g}^a}P_L s^\alpha )\, ,
\end{array}
\end{equation}

\item one gluino and one neutralino:
\begin{equation} \label{neutr:glu}
\begin{array}{lll}
Q_1^{g\chi} &=& \sqrt{\frac{1}{\alpha_w}}\, \frac{1}{2} \left[ t^{b}_{\beta \alpha} (\overline{d}^\beta P_R \widetilde{g}^b)(\overline{\widetilde{W}^3}P_L s^\alpha )  + (\widetilde{g}^b \leftrightarrow \widetilde{W}^3) \right]\\
Q_2^{g\chi} &=& \sqrt{\frac{1}{\alpha_w}}\, \frac{1}{2} \left[ t^{b}_{\beta \alpha} (\overline{d}^\beta P_R \widetilde{g}^b)(\overline{\widetilde{B}}P_L s^\alpha )+ (\widetilde{g}^b \leftrightarrow \widetilde{B}) \right] \, ,
\end{array}
\end{equation}

\item two neutralinos or two charginos:
\begin{equation} \label{neutr:charg}
\begin{array}{lll}
Q_1^{\chi} &=&{\frac{1}{\alpha_w}}\,(\overline{d}^\alpha P_R \widetilde{W}^3)(\overline{\widetilde{W}^3}P_L s^\alpha )\\
Q_2^{\chi} &=&{\frac{1}{\alpha_w}}\,\frac{1}{2} \left[ (\overline{d}^\alpha P_R \widetilde{W}^3)(\overline{\widetilde{B}}P_L s^\alpha ) + (\widetilde{W}^3 \leftrightarrow \widetilde{B})  \right]\\
Q_3^{\chi} &=& {\frac{1}{\alpha_w}}\,(\overline{d}^\alpha P_R \widetilde{B})(\overline{\widetilde{B}}P_L s^\alpha )\\
Q_4^{\chi} &=& {\frac{1}{\alpha_w}}\,(\overline{d}^\alpha P_R \widetilde{W}^+)(\overline{\widetilde{W}^+}P_L s^\alpha )\, .
\end{array}
\end{equation}
\end{itemize}

In the case of $Q_1^{g\chi}$, $Q_2^{g\chi}$ and $Q_2^{\chi}$ the antisymmetric combination of the two terms is generated at high energy with null coefficient and then receives contributions which are not relevant for us, since we are just interested in $C_1$ up to order $\xi_h \xi_\ell$ and $\eta_h \eta_\ell$.

Notice the rescaling factors in the definition of the operators (\ref{neutr:glu}) and (\ref{neutr:charg}), analogous to what 
done in \cite{Misiak:1992bc}\cite{Buras:1994qa} .
In the following we will keep $\alpha_w$ constant and the ADM at lowest order in $\alpha_s$.

The appropriate effective Lagrangian  to work with is:
\begin{equation}
\mathcal{L}^{eff} = C_1 Q_1 + \sum_{i=1}^3 C_{i}^g Q^g_i  +  \sum_{i=1}^2 C_{i}^{g\chi} Q^{g\chi}_i  +  \sum_{i=1}^4 C_{i}^{\chi} Q^{\chi}_i \, .
\end{equation}
It is convenient to define the scale-dependent 10-component vector
\begin{equation}
{\bf C}^T = (C_1, \hat{C}_g^T, \hat{C}_{g\chi}^T, \hat{C}_{\chi}^T); \quad \left(\hat{C}_g\right)_i = C^g_i \, , \, \left(\hat{C}_{g\chi}\right)_i = C^{g\chi}_i \, , \, \left(\hat{C}_{\chi}\right)_i = C^{\chi}_i
\end{equation}
satisfying an appropriate initial condition at $\mu = m_h$, and a Renormalization Group Equation (RGE):
\begin{equation}
\frac{d {\bf C}}{d \log{\mu}} = { \Gamma}^T {\bf C}.
\end{equation}
The $10\times 10$ matrix of anomalous dimensions, $\Gamma$, receives contributions both from standard gluon exchanges as from flavour-changing light-squark exchanges. Its explicit expression for a generic $SU(N)$ of colour is:
\begin{equation}
\Gamma=\frac{\alpha_s}{2\pi}\left(
\begin{array}{cccc}
\gamma _1& \xi _\ell \hat{\gamma}_{1g} & \frac{\alpha_w}{\sqrt{\alpha_s}} \xi _\ell \hat{\gamma}_{1g\chi} &  \frac{\alpha_w^2}{\alpha_s} \xi _\ell \hat{\gamma}_{1\chi}\\
\xi _\ell {\hat{\gamma}_{g1}}^{\, T} &  \hat{\gamma}_{gg} & 0 & 0 \\
 \frac{\xi _\ell}{\sqrt{\alpha_s}} {\hat{\gamma}_{g\chi 1}}^{\, T} & 0 & {\gamma}_{g\chi} \cdot 1_{2\times 2} & 0 \\
 \frac{\xi _\ell}{\alpha_s} {\hat{\gamma}_{\chi 1}}^{\, T} & 0 & 0 & {\gamma}_{\chi\chi}\cdot 1_{4\times 4}
\end{array}
\right),
\end{equation}
where $\gamma_1 = 3\frac{N-1}{N}$ is the standard anomalous dimension of $Q_1$, while:
\begin{equation}
\hat{\gamma}_{g1}  =  \left(\frac{N^2-1}{4 N}, \frac{(N^2-4)(N-1)}{8N^2},  \frac{N-1}{8}\right)
\label{eq:gammas}
\end{equation}
\begin{equation}
\hat{\gamma}_{g\chi 1}  =  \left( -\frac{N-1}{8 N},\frac{N-1}{24 N}t_W \right)
\label{eq:gammas2}
\end{equation}
\begin{equation}
\hat{\gamma}_{\chi 1}  =  \left(\frac{1}{8} , -\frac{1}{24}t_W , \frac{1}{72} t_W^2 ,\frac{\eta_\ell}{\xi_\ell } \frac{1}{4} \right)
\label{eq:gammas3}
\end{equation}
\begin{equation}
\hat{\gamma}_{gg}=
\left(
\begin{array}{ccc}
\frac{n_\ell}{4} & 0 & -6 \\
0 & -\frac{3N}{2} + \frac{n_\ell}{4} & -\frac{3N}{2} + \frac{6}{N} \\
-3 & -\frac{3N}{2} & -\frac{3}{2}N + \frac{n_\ell}{4}
\end{array}
\right)
\end{equation}
\begin{equation}
{\gamma}_{g\chi} =  -\frac{3N}{2} + \frac{n_\ell}{8}
   \quad , \quad {\gamma}_{\chi \chi} = 0     \, 
\end{equation}
where $n_\ell$ is the number of light squarks ($n_\ell=3$ if $\tilde{t}_L \, , \, \tilde{t}_R \, , \, \tilde{b}_L$ as we assume; $n_\ell=4$ if also $\tilde{b}_R$).

An important observation is in order. As already stated we are interested only in the expression for $C_1(m_\ell)$ at the light scale up to order 
$\xi_h \xi_\ell$ and $\eta_h \eta_\ell$ \footnote{If this is not true one should consider many other terms, including the $\Delta F = 0$ 
operators with four
gaugino external legs and $\hat{\gamma}_{1 g \chi}$,
$\hat{\gamma}_{1 \chi}$ and $\hat{\gamma}_{1 g}$.}.
For this reason we do not care about the terms $\hat{\gamma}_{1g}$, $\hat{\gamma}_{1g\chi}$ and $\hat{\gamma}_{1\chi}$, which introduce subleading 
corrections. 
Notice that, besides all terms of the type $\alpha_s^{2+n} \log^{1+n}$, we are also obtaining {\it some} of the terms of the type $\alpha_w \alpha_s^{1+n}  \log^{1+n}$ 
and $\alpha_w^2 \alpha_s^n \log^{1+n}$. In fact we obtain {\it all} such terms in the limit in which $\alpha_w$ is left constant and we do not dress the 
various operators with genuine weak interactions.
This means in particular that this procedure includes the relevant $\alpha_w^2 \alpha_s^n \log^{1+n}$ terms in the limit $\xi_\ell \xi_h \ll \eta_\ell \eta_h$, in which they could dominate over the other contributions.

It is also clear that all the entries of $\Gamma$ which are due to gluon exchange do not depend on the assumptions on the flavour structure, since 
flavour changing couplings enters only in the non-diagonal part of $\Gamma$ as well as in the initial conditions.

\section{Improved evolution from $m_h$ to $m_\ell$}   \label{sect:evol}

To obtain $C_1(m_\ell)$ one has first to evolve to the scale $\mu$ the coefficients of the $\Delta F=1$ operators, which is readily done by diagonalizing the $3\times 3$ matrix $\hat{\gamma}_{gg}$, via $\hat{\gamma}_{gg}^T = A \hat{\gamma}_{gg}^D A^{-1}$.
In terms of $A$ and of the diagonal matrix $\hat{\gamma}_{gg}^D$, one has:
\begin{eqnarray}
\hat{C}_g (\mu) &=&  A \left( \frac{\alpha_s(\mu)}{\alpha_s(m_h)}\right)^{\hat{\gamma}_{gg}^{D} /b_0} A^{-1} \hat{C}_g (m_h),
\label{eq:solDF1} \\
\hat{C}_{g\chi} (\mu) &=&   \left( \frac{\alpha_s(\mu)}{\alpha_s(m_h)}\right)^{{\gamma}_{g\chi} /b_0}  \hat{C}_{g\chi} (m_h) \\
\hat{C}_{\chi} (\mu) &=&   \left( \frac{\alpha_s(\mu)}{\alpha_s(m_h)}\right)^{{\gamma}_{\chi\chi} /b_0} \hat{C}_{\chi} (m_h) \label{eq:solChi}
\end{eqnarray}
where $b_0/2\pi$ is the one loop coefficient of the beta-function for $\alpha_s$.

The RGE for $C_1$ has now the form:
\begin{equation}
\frac{dC_1}{d\log \mu}=  \frac{\alpha_s}{2\pi}\gamma_1 C_1 + \frac{\alpha_s}{2\pi} \xi_\ell  \hat{\gamma}_{g1}^T \hat{C}_g + 
\frac{\sqrt{\alpha_s}}{2\pi} \xi_\ell  \hat{\gamma}_{g\chi 1}^T \hat{C}_{\chi g} + 
\frac{1}{2\pi} \xi_\ell  \hat{\gamma}_{\chi 1}^T \hat{C}_\chi,
\label{eq:RGEDF2}
\end{equation}
with $\hat{C}_g ,~ \hat{C}_{\chi g} ,~ \hat{C}_\chi$ given in eqs. (\ref{eq:solDF1})-(\ref{eq:solChi}).
The analytic solution of this RGE in our approximations is:
\begin{eqnarray}
C_1 (m_\ell) &=& \left( \frac{\alpha_s(m_\ell)}{\alpha_s(m_h)}\right)^{\gamma_1/b_0}C_1 (m_h)+ \xi_\ell   \hat{\gamma}_{g1}  A B_D A^{-1}\hat{C}_g (m_h) \label{eq:solDF2} \\
&+&  \frac{\xi_\ell  \hat{\gamma}_{g\chi 1} \hat{C}_{g\chi}(m_h)}{\gamma_{g\chi}-\gamma_1 - \frac{1}{2}b_0} \left[  
\frac{1}{\sqrt{\alpha_s(m_\ell)}} \left( \frac{\alpha_s(m_\ell)}{\alpha_s(m_h)}\right)^{\gamma_{g\chi}/b_0} -   
\frac{1}{\sqrt{\alpha_s(m_h)}}\left(  \frac{\alpha_s(m_\ell)}{\alpha_s(m_h)}\right)^{\gamma_1/b_0} \right] \nonumber \\
&+&  \frac{\xi_\ell  \hat{\gamma}_{\chi1} \hat{C}_{\chi}(m_h)}{\gamma_{\chi\chi}-\gamma_1 - b_0} \left[  
\frac{1}{\alpha_s(m_\ell)}\left( \frac{\alpha_s(m_\ell)}{\alpha_s(m_h)}\right)^{\gamma_{\chi\chi}/b_0}  - 
\frac{1}{\alpha_s(m_h)}  \left(  \frac{\alpha_s(m_\ell)}{\alpha_s(m_h)}\right)^{\gamma_1/b_0} \right] \nonumber
\end{eqnarray}
with the matrix elements of the diagonal matrix $B_D$ given by:
\begin{equation}
(B_D)_{kk}= \frac{1}{\gamma_{k}-\gamma_1} \left[ \left( \frac{\alpha_s(m_\ell)}{\alpha_s(m_h)}\right)^{\gamma_k/b_0}-\left(  \frac{\alpha_s(m_\ell)}{\alpha_s(m_h)}\right)^{\gamma_1/b_0}\right]\; ,~\gamma_k = (\hat{\gamma}^D_{gg})_{kk}.
\label{eq:Gamma}
\end{equation}
The first term on the right-hand-side of (\ref{eq:solDF2}) corresponds to the standard rescaling of $Q_1$, whereas the other terms,
proportional to $\xi_\ell $ are the QCD corrected contribution appearing at lowest order in $\mathcal{L}^{\Delta F =2}_{h\ell}$,
eqs (\ref{eq:L123}) and (\ref{eq:naive}).

The relevant initial conditions at the heavy scale are (with $N=3$):
$$
C_1 =\xi_h \xi_\ell \frac{1 }{m_h^2} \left[- \frac{37}{36}\alpha_s^2 + \alpha_w^2 \left( \frac{1}{4}\frac{\eta_h \eta_\ell}{\xi_h \xi_\ell}- \frac{5}{2}
\left( \frac{1}{4}\frac{\eta_h \eta_\ell}{\xi_h \xi_\ell} +\frac{2}{3} \frac{\alpha_s}{\alpha_w}\mathcal{R}+\mathcal{R}^2 \right) \right) \right]
$$
\begin{equation}
\hat{C}_g^T = -4\pi \alpha_s\xi_h \frac{1}{m_h^2} \left( \frac{1}{3}, 1, 1 \right)\, ,  ~~~~~ 
\hat{C}_{g\chi}^T = -4\pi \alpha_w \sqrt{\alpha_s} \xi_h \frac{1}{m_h^2} \left( - 2, \frac{2}{3}t_W \right)
\end{equation}
$$
\hat{C}_{\chi}^T = -2\pi \alpha_w^2\xi_h \frac{1}{m_h^2} \left( 1, -\frac{2}{3}t_W, \frac{1}{9}t_W^2 , 2 \frac{\eta_h}{\xi_h} \right)
$$
It is easy to see that expanding the full result up to order $\alpha_s^2 \, , \, \alpha_w^2 \, , \, \alpha_s \alpha_w$ one obtains exactly the logarithms appearing in (\ref{eq:naive}).
$C_1(m_\ell)$ can then be evolved down to the GeV scale in a standard way, properly accounting for the different thresholds one encounters in the beta-function coefficient.
\\

Let us finally say something about the relative size of these corrections.
Consider the illustrative case $m_\ell = 500$ GeV and $m_h=$ 20 TeV \footnote{Notice that this value is not necessarily unnatural, see \cite{Barbieri:2010pd}.} with $\xi_\ell \xi_h = \eta_\ell \eta_h$.
What one finds is that the relative correction to the Wilson coefficient of $Q_1$, due to light-heavy exchange only, is respectively about $45 \%$ for the terms proportional to $\alpha_s^2$, $3 \%$ for those proportional to $\alpha_s \alpha_w$ and $2 \%$ for the terms proportional to $\alpha_w^2$. This is the correction with respect to using (\ref{eq:naive}) as initial condition at $m_h$ and then naively evolve the coefficient with $\gamma_{1}$ only (as usually done in the literature \cite{Bagger:1997gg}-\cite{Contino:1998nw})\footnote{This result justifies the approximation used in \cite{Barbieri:2010ar} where only the corrections to the Wilson coefficient proportional to $\alpha_s^2$ are considered.}.
If, on the contrary, the amplitude (\ref{eq:naive}) were used as initial condition at the scale $m_\ell$, then the relative errors would be respectively $35 \%$ (from $\alpha_s^2$), $1 \%$ ($\alpha_s \alpha_w$) and $1 \%$ ($\alpha_w^2$).
Notice that, for moderate hierarchy, the relative impact of this correction is of the same order of what one typically obtains by computing the exact NLO corrections and then going to the hierarchical limit \cite{Virto:2009wm}. In fact in that way one gets the correct result up to order $\alpha_s^4 \log^3$, together with terms $\alpha_s^4 \log^2$ that are beyond our approximation. However if the logarithms are really large then it is necessary to completely resum them, as we did here for the LO result.

The dominance of the corrections to the $\alpha^2_s$ term reflects the fact that at one loop order the gluino-gluino exchanges give by themselves the $90-95\%$ of the total. This is the case if $\xi_\ell \xi_h = \eta_\ell \eta_h$. Notice however that, for $\xi_\ell \xi_h < \eta_\ell \eta_h$, the proper corrections to the term $\alpha_w^2 \log$ can become important.



\section*{Acknowledgments}

We thank Riccardo Barbieri for important suggestions and Gino Isidori for useful comments.
We also thank Raffaele Tito D'Agnolo and Dmitry Zhuridov.
This work is supported in part by the European Programme ``Unification in the LHC Era",  contract PITN-GA-2009-237920 (UNILHC).

\end{document}